\documentstyle[12pt]{article}
 \let\msk=\medskip \let\bsk=\bigskip
\let\qd=\quad  
\let\a=\alpha \let\be=\beta \let\g=\gamma \let\de=\delta
\let\ep=\varepsilon  \let\h=\eta 
  \let\la=\lambda \let\m=\mu
\let\n=\nu  \let\r=\rho \let\si=\sigma
 \let\ps=\psi
\let\ph=\varphi   \let\Ps=\Psi
  
\let\La=\Lambda  

\def\L{\de}
\def\0#1#2{\frac{#1}{#2}} \def\s0#1#2{\mbox{\small{$\frac{#1}{#2}$}}}
\def\2{{\times}} \def\3{\vec }
\def\5{\bar }  \def\6{\partial } \def\7{\hat } \def\4{\tilde }
 \def\lb{\left(} \def\rb{\right)}
\def\lra{ \leftrightarrow } \let\LRA=\Leftrightarrow
\let\then=\Rightarrow 
\def\sl2c{$sl(2,{\bf C})$}
\def\SL2c{$SL(2,{\bf C})$}

\let\nn=\nonumber
\def\bea{\begin{eqnarray}} \def\eea{\end{eqnarray}}
\def\beann{\begin{eqnarray*}} \def\eeann{\end{eqnarray*}}
\def\beq{\begin{equation}} \def\eeq{\end{equation}}
\def\ba{\begin{array}} \def\ea{\end{array}}
\def\ben{\begin{enumerate}} \def\een{\end{enumerate}}

 \def\cb{{\cal B}} \def\cg{{\cal G}}
\def\ca{{\cal A}}  \def\cl{{\cal L}}
  \def\cM{{\cal M}}
  
\def\cW{{\cal W}}

\def\T#1#2#3{T_{#1#2}{}^{#3}}
\def\F#1#2#3{{F_{#1#2}}^{#3}}
\def\f#1#2#3{{f_{#1#2}}^{#3}}

\def\A#1#2{{A_{#1}}^{#2}}
\def\tA#1#2{\4A_{#1}{}^{#2}}

\newcommand{\mysection}[1]{\section{#1}\setcounter{equation}{0}}
\def\Gl#1{(\ref{#1})}

\begin{document}
{\pagestyle{empty}
\hfill NIKHEF-H 93-23

\vspace*{1cm}

\begin{center}{\LARGE {\bf
B\"acklund Transformations and\vspace{.5cm}\\
Zero-Curvature Representations of \vspace{.5cm}\\
Systems of\vspace{.5cm}\\
Partial Differential Equations
}}\vspace{1cm}

{\renewcommand{\thefootnote}{\fnsymbol{footnote}}
{\Large Friedemann Brandt}\footnote{Supported by Deutsche
Forschungsgemeinschaft}}
\setcounter{footnote}{0}\vspace{.5cm}

NIKHEF-H, P.O.~Box 41882,
1009 DB Amsterdam, The Netherlands\end{center}\vspace{.5cm}

\begin{abstract}
It is shown that B\"acklund transformations (BTs) and
zero-curvature representations (ZCRs) of systems of partial differential
equations (PDEs) are closely related. The connection is established
by nonlinear representations of the symmetry group underlying the ZCR
which induce gauge transformations relating different BTs.
This connection is used to construct BTs from ZCRs (and vice versa).
Furthermore a procedure is outlined which allows a systematic search
for ZCRs of a given system of PDEs.
\end{abstract}
\vspace{1cm}

\begin{center} Accepted for publication in J.Math.Phys.\end{center}
\newpage}\setcounter{page}{1}

\begin{center}{\Large {\bf B\"acklund Transformations and
Zero-Curvature Representations of Systems of
Partial Differential Equations}}\vspace{.5cm}

{\renewcommand{\thefootnote}{\fnsymbol{footnote}}
{\Large Friedemann Brandt}\footnote{Supported by
Deutsche Forschungsgemeinschaft}}
\setcounter{footnote}{0}\vspace{.5cm}

NIKHEF-H, P.O.~Box 41882,
1009 DB Amsterdam, The Netherlands\end{center}

\begin{abstract}
It is shown that B\"acklund transformations (BTs) and
zero-curvature representations (ZCRs) of systems of partial differential
equations (PDEs) are closely related. The connection is established
by nonlinear representations of the symmetry group underlying the ZCR
which induce gauge transformations relating different BTs.
This connection is used to construct BTs from ZCRs (and vice versa).
Furthermore a procedure is outlined which allows a systematic search
for ZCRs of a given system of PDEs.
\end{abstract}

\mysection{Introduction}

During the past three decades so-called integrable systems of
nonlinear partial differential equations (PDEs) have attracted much
interest both in physics and in mathematics.
This interest is owing to the numerous
applications which such equations have in many different branches
of physics and to the rich mathematical structures which showed up
behind them. Such structures are the existence of
Lax pairs, Miura maps,
B\"acklund transformations, infinitely many local conservation laws,
(bi-)Hamiltonian structures and the applicability
of inverse scattering methods.

Celebrated examples of integrable nonlinear
PDEs which have some or all of these
remarkable properties are the Korteweg-de Vries
(KdV) equation, the modified Korteweg-de Vries (mKdV) equation, the
Sine-Gordon equation, the Liouville equation and the nonlinear Schr\"odinger
equation. Meanwhile one knows
infinitely many systems of nonlinear PDEs with these properties.
A famous example of an
infinite set of nonlinear integrable PDEs is given by the KdV hierarchy
which in fact itself is just one member of an infinite set of related
hierarchies \cite{ds}.

However our knowledge about structures
related with integrability as those mentioned above
is still incomplete and in many respects gives the impression of
an accumulation of examples and methods whose deeper
origin, connection or range of applicability are not completely understood
yet. In particular methods are
lacking which allow to test a given system of PDEs for integrability and
to find the related mathematical structures systematically.

A key for progress in this field may be provided by an improved
understanding of a particular property which many, if not all
known integrable systems of PDEs have.
This property is the existence of a
zero-curvature representation (ZCR). A ZCR of a system of PDEs for
functions $u_a$ consists of a set of Lie algebra valued `gauge fields'
$\ca_\m(x,u,\6u,\6^2u,\ldots)$ constructed of
the $u_a$, their partial derivatives
and the coordinates $x^\m$ of the underlying manifold
such that the vanishing of the field strengths
$\6_\m\ca_\n-\6_\n\ca_\m-[\ca_\m,\ca_\n]$
encodes the system of PDEs for the $u_a$ (this will be made more
precise later).

A ZCR may be regarded as a master property of an integrable system of PDEs
since often other important properties of the system and methods
for its investigation can be derived from it.
For instance the inverse scattering
method \cite{akns} for solving special systems of
nonlinear PDEs is based
on ZCRs of these PDEs \cite{cramp}, and the Lax representation
\cite{lax} of a
system of nonlinear PDEs essentially is nothing but a ZCR \cite{ds}
(in fact a ZCR of a system of PDEs may be regarded
as a generalized Lax representation). In some cases a ZCR itself may have
interesting physical interpretation, see e.g. \cite{das3}.

Another remarkable property of many integrable systems of PDEs is the
existence of B\"acklund transformations (BTs). The latter
have proved to be particularly useful for the construction of
solutions of systems of nonlinear PDEs.
For instance a BT may relate a system of
PDEs to a simpler one whose solutions can be used to
construct solutions of the more complicated system by means of the BT.
Or a BT may
relate different solutions of the same system. Then it is called an
auto-BT (or self-BT) for this system and may be used to construct
complicated solutions
of the system from simpler solutions by an algebraic method based on
the so-called permutability of auto-BTs
\cite{das1,dra,eis}. Other useful applications
may arise if a BT contains a Miura map whose exponent has been found in
\cite{miu} and has been used to prove the
existence of an infinite set of local conservation laws for the KdV equation
and to construct these conservation laws explicitly \cite{miu2}.
Miura maps have interesting applications especially
for Hamiltonian systems of evolution equations
if they relate different Hamiltonian structures \cite{af}. Such
`Hamiltonian Miura maps' may be useful even for the quantization of
conformal field theories \cite{fl}. This application originates in
the connection of conformal field theories with (bi-)Hamiltonian systems
of PDEs via their (second) Hamiltonian Poisson bracket structure \cite{mat}
which provides a realization of so-called classical $\cW$--algebras
whose investigation was initiated by \cite{zamo}.

Interrelations between BTs and ZCRs have been noticed by several authors.
For instance a connection of the pioneering work \cite{we2} with
zero-curvature conditions has been pointed out already in \cite{her1}.
The present paper works out
close relationships of BTs and ZCRs. Namely
it turns out that a BT of a certain (rather general) type for a
given system of PDEs induces a corresponding ZCR of this system
and, conversely, that a ZCR of a system can be used to construct
BTs for it.

The connection between ZCRs and BTs is established
by means of gauge transformations which relate different BTs and
represent the gauge group underlying the ZCR
in general nonlinearly on an infinite dimensional function space.
In particular this allows to define gauge equivalence of BTs.
A second, more technical ingredient used to relate BTs with ZCRs
is an approach which can be formalized
using the jet bundle theory. The latter provides a suitable mathematical
framework for an investigation of algebraic aspects of PDEs in general,
see e.g. \cite{olv}. We shall only need some very elementary
ideas underlying the jet-bundle theory. A systematic and more formal
application of this theory to BTs can be found e.g. in \cite{pir}.

The paper is organized as follows. In section \ref{bt}
the type of BTs is defined which are considered in this paper and the
basic notation is introduced. Furthermore some celebrated examples of
BTs are recalled
which are used later for exemplifications. In section \ref{zcr}
the connection between BTs and ZCRs is worked out. Section \ref{gauge}
introduces the above-mentioned gauge transformations.
In section \ref{diff} nonlinear representations
of Lie groups are discussed which are supposed to be particular
interesting in this context
since they forge links to inverse scattering techniques.
The general procedure for the construction
of BTs from given ZCRs is exemplified in section \ref{Lax} for
the generalized KdV hierarchies. Finally
in section \ref{meth} a method is outlined and exemplified
which allows a systematic search for a ZCR of a given system of PDEs.

\mysection{B\"acklund transformations}\label{bt}

Let me first introduce some notation. The BTs considered later will
generally relate two sets of functions $\{u_a\}$, $\{v_i\}$
\beq u_a=u_a(x),\qd a=1,\ldots,N_u,\qd v_i=v_i(x),\qd i=1,\ldots,N_v
\label{b1}\eeq
where $N_u$ and $N_v$ are not necessarily equal. The
argument $x$ of these functions refers to some coordinate system
\beq x^\m,\qd \m=1,\ldots,D\label{b2}\eeq
of the underlying basis manifold whose
dimension $D$ will not be fixed in the general case. However since
all examples will refer to the case $D=2$ we shall also
use the customary notation
\beq D=2:\qd t=x^1,\qd x=x^2\label{b2a}\eeq
hoping this will not cause confusion with the collective notation $x$
for arguments of functions in the general ($D$-dimensional) case as in
\Gl{b1}.

$[u]$ denotes collectively
the functions $u_a$ and their partial derivatives,
i.e. the whole set of variables
\beq \left\{\0 {\6^nu_a}{\6x^{\m_1}\ldots\6x^{\m_n}}:\qd a=1,\ldots,N_u,\qd
                                                    n=0,1,2,\ldots\right\}.
\label{b3}\eeq
In fact a suitable subset of \Gl{b3}
will mostly be regarded as a set of independent variables
instead of regarding its elements as functions of the coordinates
$x^\m$. This approach is formalized in the above-mentioned jet bundle theory.

According to this remark it should be clear that it will be important
what are the relevant variables in a present context. This will be
indicated mostly by the arguments of a function which usually are written
omitting any indices.
Thus $f(v,x,[u])$ will in the general case
denote a function of the variables $x^\m$, $v_i$ and the elements of \Gl{b3}.

Let us now define the BTs which we shall deal with.
A system of $D$-dimensional partial differential equations of the form
\beq  \0 {\6v_i}{\6x^\m}=A_{\m{i}}(v,x,[u])
\label{1}\eeq
will be called a BT if its integrability conditions
hold owing to a system of PDEs satisfied by the
$u_a$ which we denote by
\beq P_A(x,[u])=0,\qd A=1,\ldots,N_P.\label{b4}\eeq
We do not insist on $N_P=N_u$ which is of course the most important case
(however also `overdetermined' systems of PDEs with $N_P> N_u$
may be interesting, see e.g. \cite{cr,das2}).
Regarding $A_{\m{i}}(v,x,[u])$ as functions $B_{\m i}(x)$ of the
coordinates $x^\m$, the integrability conditions for \Gl1 read of course
\beq \0 {\6B_{\m i}(x)}{\6x^\n}-\0 {\6B_{\n i}(x)}{\6x^\m}=0,\qd
B_{\m i}(x)=A_{\m i}(v,x(x),[u(x)])
\label{2}\eeq
but this is not a useful point of view in the present context since it
does not make contact with \Gl{b4}. Therefore we shall introduce
more useful versions of the integrability conditions for \Gl1 in section
\ref{zcr}.
Notice that by assumption $A_{\m i}(v,x,[u])$ does not
depend on derivatives of $v$, i.e. we consider only
systems \Gl1 of first order in the derivatives of $v$. However
since higher order systems can be rewritten in first
order form, \Gl1 is more general than it may appear at first sight.

Furthermore it is stressed that generally
we do not require that \Gl1 implies a system of PDEs
\beq Q_B(x,[v])=0,\qd B=1,\ldots,N_Q\label{b5}\eeq
for the $v_i$ as well. This should be kept in mind since
the definition of BTs of the form \Gl1 often is restricted to the cases where
\Gl1 and its integrability conditions imply both \Gl{b4} and \Gl{b5}.
In particular \Gl1 is called an auto-BT if these systems are equal ($P=Q$).
\bsk

\noindent{\it Examples:}\\
Let us conclude this section with some celebrated examples of BTs
for the simplest case $N_u=N_v=1$, $D=2$. We shall use
the notation \Gl{b2a} and denote differentiations with respect to
$x$ or $t$ by subscripts ($v_x=\6v/\6x$ etc.). Furthermore we use
$v=v_1$ and $u=u_1$ resp. $w=u_1$, and in all examples
$\la$ denotes an arbitrary constant (spectral parameter) and \Gl1 and
its integrability conditions lead to decoupled PDEs for
$u$ (resp. $w$) and $v$ denoted by $P([u])=0$ (resp. $P([w])=0$) and
$Q([v])=0$.
\ben
\item[a)] The classical example for a BT is the
auto-BT for the Sine-Gordon equation which
in light-cone coordinates reads
\bea & & v_x=-u_x+\la \sin\0 {u-v}2,\qd v_t=u_t-\0 4\la \sin\0 {u+v}2,
\label{bsg}\\
& & P([u])=Q([u])=u_{xt}-\sin u=0.\label{sg}\eea
\item[b)] A BT which relates the Liouville equation
and the two-dimensional d'Alembert
equation is in light-cone coordinates given by
\bea & & v_x=u_x+\la \exp\0 {u+v}2,\qd v_t=-u_t-\0 2\la \exp\0 {u-v}2,
\label{bl}\\
& & P([u])=u_{xt}-\exp u=0,\qd Q([v])=v_{xt}=0.\label{l}\eea
\item[c)] A BT which relates the KdV equation and the ($\la$-dependent) mKdV
equation is given by \cite{lamb}
\bea & & v_x=\la+u+v^2,\nn\\
& & v_t=-u_{xx}-2u^2+2\la u+4\la^2-2u_xv+v^2(4\la-2u),
\label{bmiu}\\
& & P([u])=u_t+u_{xxx}+6uu_x=0,\label{kdv}\\
& & Q([v])=v_t+v_{xxx}-6v^2v_x-6\la v_x=0.\label{mkdv}\eea
The `space part' of the BT \Gl{bmiu} is the famous Miura map \cite{miu}.
\item[d)] An auto-BT for the `potential KdV' (pKdV) equation is given by
\cite{we1}
\bea & &\ba{l} v_x=-w_x-2\la-\s0 12(w-v)^2,\\
v_t=w_{xxx}+(w_x)^2-4\la w_x
-8\la^2+2w_{xx}(w-v)+(w_x-2\la)(w-v)^2,\ea\label{bkdv}\\
& &P([w])=Q([w])=w_t+w_{xxx}+3(w_x)^2=0.\label{pkdv}\eea
\Gl{pkdv} is called the pKdV equation since its solutions
serve as potentials for solutions of the KdV equation
($u=w_x$ solves \Gl{kdv} if $w$ solves \Gl{pkdv}).
\item[e)] An auto-BT for the `potential mKdV' (pmKdV) equation
reads \cite{lamb}
\bea  v_x&=&w_x+\la \sinh (v+w),\nn\\
   v_t&=&-w_{xxx}+2(w_x)^3-2\la^2w_x-2\la w_{xx}\cosh(v+w)\nn\\ & &
          +\la\lb 2(w_x)^2-\la^2\rb\sinh(v+w),\label{bmkdv}\\
 P([w])&=&Q([w])=w_t+w_{xxx}-2(w_x)^3=0.\label{pmkdv}\eea
\een
\noindent{\it Remarks:}

(i) The BTs \Gl{bmiu} and \Gl{bkdv} can both be obtained from
a `mother-BT' which we shall construct in section \ref{meth}
(see eq. \Gl{pp8}) and are gauge equivalent, see section \ref{gauge}.

(ii) The auto-BTs \Gl{bkdv} and \Gl{bmkdv} are usually written
in forms which have a symmetry under exchange of $v$ and $w$
rather than in the form \Gl1. For instance using \Gl{pmkdv}
and the `space part' of
\Gl{bmkdv} one can write its `time part' in the form
\[ v_t-w_t=
\la\lb(v_x)^2+(w_x)^2\rb\sinh(v+w)-\la(v_{xx}+w_{xx})\cosh(v+w)\]
which, as the `space part',
is invariant under $v\leftrightarrow w$, $\la\rightarrow -\la$
(the same symmetry occurs in \Gl{bsg}).

\mysection{The connection between B\"acklund transformations
and zero-\-cur\-va\-ture representations}\label{zcr}

In order to establish the connection between BTs of the form \Gl1
and a ZCR of a system \Gl{b4} we first
write the integrability condition \Gl2 in a more useful form by regarding
$A_{\m i}$ not as a function of $x^\m$ as in \Gl2 but as a function of
the variables $x^\m$, $v_i$ and \Gl{b3} as in \Gl1.
Taking advantage of \Gl1 we represent
the partial derivatives on these variables by the operators
\beq D_\m=\6_\m+A_{\m i}(v,x,[u])\, \0 \6{\6v_i}\label{4A}\eeq
where Einstein's summation convention is used (summation over $i$) and
the piece $\6_\m$ acts nontrivially only on
the variables \Gl{b3} and on the $x^\m$:
\beq\6_\m\, \0 {\6^nu_a}{\6x^{\m_1}\ldots\6x^{\m_n}}
=\0 {\6^{n+1}u_a}{\6x^\m\6x^{\m_1}\ldots\6x^{\m_n}}
,\qd \6_\m x^\n=\de_\m^\n,
\qd \6_\m v_i=0.\label{5}\eeq
On functions $f(v,x,[u])$ we of course define $D_\m$
as first order differential
operator satisfying the product rule
\beq D_\m(XY)=(D_\m X)Y+X(D_\m Y).\label{6C}\eeq
We can now easily calculate the commutator
of two `partial derivatives':
\bea [D_\m,D_\n]&=&F_{\m\n i}(v,x,[u])\0 \6{\6 v_i}\, ,\label{F1}\\
 F_{\m\n i}(v,x,[u])&=&
        D_\m A_{\n i}(v,x,[u])-D_\n A_{\m i}(v,x,[u]).\label{F2}\eea
Requiring that the commutator \Gl{F1} vanishes
expresses the integrability condition for \Gl1 since they
read in terms of the variables $x,v,[u]$:
\beq 0=[D_\m,D_\n]\, v_i=F_{\m\n i}(v,x,[u]).\label{z3A}\eeq
\Gl{z3A} has already the form of a zero-curvature condition
imposed on the `field strengths' $F_{\m\n i}$ of the `gauge fields'
$A_{\m i}(v,x,[u])$. Notice however that \Gl{z3A} involves the $v_i$.
According to the previous section we assume that \Gl{z3A}
holds by virtue of a system
of PDEs \Gl{b4} for the $u_a$ which does {\it not} involve the $v_i$.
This will be used now to extract from \Gl1 a ZCR of \Gl{b4}.
To this end we decompose the right-hand sides of \Gl1 according to
\beq A_{\m i}(v,x,[u])=\A \m{I}(x,[u])\, R_{I i}(v)\label{3A}\eeq
where again summation over $I$ is understood and
$\{R_{I i}\}$ denotes a set of linearly independent functions.
The differential operators
\beq \L_I=-R_{I i}(v)\, \0 \6{\6 v_i}\label{6A}\eeq
span a Lie algebra $\cg$ whose structure constants are
denoted by $\f IJK$:
\beq [\L_I,\L_J]=\f IJK\, \L_K.\label{7A}\eeq
If possible one of course
chooses the $R_{I i}$ such that $\cg$ is finite.
The operators $D_\m$ now take the familiar form of covariant derivatives
in Yang-Mills theories
\beq D_\m=\6_\m-\A \m{I}(x,[u])\, \L_I                  \label{4}\eeq
which suggests to call the $\A \m{I}(x,[u])$ gauge fields.
This terminology will be justified in the next section.
\Gl{F1} now takes the form
\beq [D_\m,D_\n]=-\F \m\n{I}(x,[u])\, \L_I\label{F1a}\eeq
where the $\F \m\n{I}(x,[u])$ are the field strengths constructed
of the gauge fields $\A \m{I}(x,[u])$ according to
\beq \F \m\n{I}(x,[u])=
\6_\m \A \n{I}-\6_\n \A \m{I}-\f JKI\A \m{J}\A \n{K},
\qd \A \m{I}=\A \m{I}(x,[u]).\label{F2a}\eeq
The $\F \m\n{I}$ are just the coefficients occurring in the decomposition
of the $F_{\m\n i}$ analogously to \Gl{3A}:
\beq F_{\m\n i}(v,x,[u])=\F \m\n{I}(x,[u])\, R_{I i}(v).\label{z1}\eeq
Since by assumption the $R_{I i}$ are linearly independent and
\Gl{z3A} holds by virtue of \Gl{b4} the latter implies
\beq \F \m\n{I}(x,[u])=0.\label{z3}\eeq
More precisely this requires
\beq \F \m\n{I}=\sum_{n}r_{\m\n}{}^{I A \r_1\ldots\r_n}
(x,[u])\, \6_{\r_1}\ldots\6_{\r_n}\, P_A(x,[u])
\label{12}\eeq
with generally nontrivial functions $r_{\m\n}{}^{I A \r_1\ldots\r_n}(x,[u])$.
Thus the $\A \m{I}(x,[u])$ indeed provide a ZCR of the system \Gl{b4}
in the sense of the definition outlined in the introduction.\bsk

\noindent{\it Examples}:\\
The BTs a)--e) listed in the previous
section provide ZCRs of the respective PDEs for $\cg=sl(2)$. The
$sl(2)$-generators are chosen such that algebra reads
\beq [\L_I,\L_J]=(I-J)\, \L_{I+J},\qd I,J=-1,0,1.\label{sl2}\eeq
A choice of the generators $\L_I=-R_I(v)\6/\6v$
which satisfies \Gl{sl2} is respectively given by
\beq
\ba{rlll}
a)  :&R_{-1}=2+2\cos\s0 v2,&R_{0}=2\sin\s0 v2 ,&R_1= 2-2\cos\s0 v2,\\
b)  :&R_{-1}= 2\exp (-\s0 v2) ,&R_{0}=2    ,&R_1=     2\exp \s0 v2,\\
c),d)  :&R_{-1}=   1   ,&R_{0}=   v         ,&R_1=        v^2     ,\\
e)  :&R_{-1}=1+\cosh v  ,&R_{0}=  \sinh v   ,&R_1=-1+ \cosh v .
\ea\label{table1}\eeq
One easily reads off the ZCR of the respective PDE
for the choice \Gl{table1} and may use it to check \Gl{12}. For instance
in the case d) one obtains
\beq\ba{lll}
\A 2{-1}= -w_x-2\la-\s0 12w^2,& \A 20=w,& \A 21=-\s0 12,\\
\multicolumn{3}{l}{\A 1{-1}=w_{xxx}+(w_x)^2-
             4\la w_x-8\la^2+2ww_{xx}+w^2(w_x-2\la),} \\
& \A 10= -2w_{xx}-2w(w_x-2\la),& \A 11=  w_x-2\la,      \\
\multicolumn{3}{l}{\F 12{-1}=-(w+\6_x)(w_t+w_{xxx}+3(w_x)^2),}\\ & \F 120=
         w_t+w_{xxx}+3(w_x)^2,& \F 121= 0.\ea\label{table2}\eeq
\noindent{\it Remark:}

It is no surprise that $sl(2)$ occurs in all examples a)--e) since
it is the unique finite Lie algebra which can occur in the case
$N_v=1$. This is easily seen as follows: Assume that $\de_I=-R_I(v)\6/\6v$
and $\de_J=-R_J(v)\6/\6v$ are two commuting elements of $\cg$. Due to
$N_v=1$ this implies
\[ R_I(v)\0 {\6R_J(v)}{\6v}-R_J(v)\0 {\6R_I(v)}{\6v}=0\ \then\
R_I(v)=c\,  R_J(v)\]
where $c$ is an arbitrary constant.
Thus $\de_I$ and $\de_J$ are linearly dependent,
i.e. the Cartan subalgebra of $\cg$ is 1-dimensional and thus $\cg$ indeed is
unique.

\mysection{Gauge transformations}\label{gauge}

The formulae of the previous section suggest the presence of
gauge transformations associated with
 the gauge fields $A_{\m i}(v,x,[u])$ resp. $\A \m{I}(x,[u])$.
Indeed they can be found. Namely consider transformations of the form
\beq \4v_i=\4v_i(v,x,[u]).
\label{13}\eeq
We call \Gl{13} a gauge transformation if the functions $\4v_i(v,x,[u])$
are local in a suitable sense\footnote{Usually
we have in mind here a definition of locality which requires
polynomial dependence on the derivatives of the $u_a$. However
the definition of locality generally should be adapted to
the particular problem \Gl{b4}.} and if \Gl{13} is
invertible in a sufficiently large
neighborhood of the point $(v_1,\ldots,v_{N_v})$ with an inverse
\beq v_i=v_i(\4v,x,[u])\label{13a}\eeq
which is local too.
The corresponding transformation of the `gauge fields' $A_{\m i}$
and their `field strengths' $F_{\m\n i}$ are given by
\bea \4A_{\m i}(\4v,x,[u])&=&\6_\m \4v_i(v,x,[u])+A_{\m j} (v,x,[u])
\, \0 {\6\4v_i(v,x,[u])}{\6v_j},                 \label{15a}\\
\4F_{\m\n i}(\4v,x,[u])&=&F_{\m\n j}(v,x,[u])\, \0 {\6\4v_i(v,x,[u])}{\6v_j}\ .
                                                           \label{15b}\eea
\Gl{15a} follows from evaluating the left-hand side of
\beq \0 {\6 \4v_i}{\6x^\m}=\4A_{\m i}(\4v,x,[u])\label{14}\eeq
using \Gl{13} and \Gl{1} (and the chain rule). \Gl{14} itself
is nothing but \Gl{1} for the transformed quantities. \Gl{15b}
follows from \Gl{15a} since the
$\4F_{\m\n i}$ are defined in terms of the $\4A_{\m i}$
analogously to \Gl{F1}:
\beq [\4D_\m,\4D_\n]=\4F_{\m\n i}(\4v,x,[u])\, \0 {\6}{\6\4v_i},\qd
\4D_\m=\6_\m+\4A_{\m i}(\4v,x,[u])\, \0 {\6}{\6\4v_i}\, .\label{4F1}\eeq
The presence of the inhomogeneous term
$\6_\m \4v_i$ in the transformation law \Gl{15a} justifies to call the
$A_{\m i}$ gauge fields. Notice that the $F_{\m\n i}$ indeed transform
covariantly and are therefore rightly called field strengths. In particular
$\4F_{\m\n i}$ vanishes if $F_{\m\n i}$ vanishes. This expresses of course
just the fact that the
integrability conditions for \Gl{14} follow from those for \Gl{1}.
BTs which are related by
gauge transformations \Gl{13} are called gauge equivalent.
Notice however that a system of PDEs \Gl{b5} for
the $v_i$ which might be implied by \Gl1
of course is not gauge invariant, i.e. the
system for the $\4v_i$ generally differs from
the system for the $v_i$. In particular it is not guaranteed that \Gl{14}
implies decoupled
systems of PDEs for $\{u_a\}$ and $\{\4v_i\}$ if \Gl1 implies decoupled
systems of PDEs for $\{u_a\}$ and $\{v_i\}$ and vice versa.

Let us now inspect how gauge transformations \Gl{13} affect the
ZCR $\A \m{I}(x,[u])$. The decomposition
of the transformed gauge fields \Gl{15a} and the
corresponding generators are denoted by
\bea  \4A_{\m i}(\4v,x,[u])&=&\tA \m{I}(x,[u])\, \4R_{I i}(\4v),\label{43A}\\
  \4\L_I&=&-\4R_{I i}(\4v)\, \0 {\6}{\6\4v_i}\, .\label{46A}\eea
For an arbitrary gauge transformation \Gl{13} it of course does not make
much sense to define individual transformations of the $R_{I i}$ and $\A \m{I}$
since in general it is
not possible to find a decomposition \Gl{43A} such that
$\4R_{I i}$ and $\tA \m{I}$ are related to their counterparts
$R_{I i}$ and $\A \m{I}$ in a simple way.
However there are two subgroups of gauge transformations \Gl{13}
which allow such decompositions and
deserve special attention.

One of these subgroups consists of those gauge transformations
which are generated by the operators $\L_I$ themselves:
\beq \4v_i(v,x,[u])=\exp \lb g^I(x,[u])\L_I\rb\, v_i\ .\label{g4}\eeq
Notice that the `parameters' $g^I$ generally depend not only on the $x^\m$ but
on the $u_a$ and their derivatives as well. They are therefore not completely
arbitrary but must be chosen such that \Gl{g4} satisfies
the above-mentioned requirements imposed by locality.
The fact that the transformations
\Gl{g4} are generated by the $\L_I$ implies that
these gauge transformations allow a decomposition \Gl{43A} given by
\beq  \4A_{\m i}(\4v,x,[u])=\tA \m{I}(x,[u])\, R_{I i}(\4v)\label{g3}\eeq
with the same functions $R_{I i}$ as in \Gl{3A}:
\beq  \4R_{I i}(\4v)=R_{I i}(\4v)\qd \forall\ I,i.\label{g3a}\eeq
In order to verify that \Gl{g3a} indeed is compatible with \Gl{g4} one
may calculate the Lie derivative of $A_{\m i}$ using \Gl{g3}. To this
end one considers `infinitesimal' gauge transformations
\[ g^I(x,[u])=\ep \, G^I(x,[u]),\qd \ep \ll 1.\]
Taking advantage of
\[ R_{I j}\0 {\6R_{J i}}{\6v_j}-R_{J j}\0 {\6R_{I i}}{\6v_j}
=-\f IJK\, R_{K i}\]
which holds by assumption according to \Gl{7A} one easily checks that
\Gl{3A}, \Gl{g4} and \Gl{g3} imply
\beq\tA \m{I}(x,[u])=\A \m{I}(x,[u])+\6_\m g^I(x,[u])+
\A \m{J}(x,[u])g^K(x,[u])\f JKI+O(\ep^2).\label{g5}\eeq
This proves on the one hand the compatibility of \Gl{g3a} and \Gl{g4}
and shows on the other hand that the $\A \m{I}$ indeed have the standard
transformation of Yang-Mills fields under gauge transformations \Gl{g4}
if the decomposition \Gl{43A} is chosen according to \Gl{g3a}.

The second above-mentioned subgroup
of gauge transformations \Gl{13} consists of those transformations
which do not depend on the $u$'s or the coordinates:
\beq \4v_i=\4v_i(v).\label{13b}\eeq
These `rigid' gauge transformations allow the decomposition
\beq \4A_{\m i}(\4v,x,[u])=\A \m{I}(x,[u])\, \4R_{I i}(\4v)\label{g1}\eeq
with the same `components' $\A \m{I}$ as in \Gl{3A} and functions
$\4R_{I i}$ which are related to the $R_{I i}$ according to
\beq\4R_{I i}(\4v)=R_{I j}(v)\, \0 {\6\4v_i}{\6v_j}.\label{g1a}\eeq
Notice that \Gl{g1a}
is a `contravariant' transformation compared to the `covariant'
transformation of the derivatives $\6/\6v_i$ under a
gauge transformation \Gl{13b}. Therefore the generators \Gl{46A}
span the same Lie algebra $\4\cg=\cg$ with the same structure
constants as the $\L_I$ for the choice \Gl{g1a}:
\beq [\L_I,\L_J]=\f IJK\L_K\qd \then\qd [\4\L_I,\4\L_J]=\f
IJK\4\L_K.\label{g2}\eeq

\noindent {\it Remark:}

It is essential to realize that the gauge fields
$\A \m{I}(x,[u])$ are not pure gauges in the space of local functions
since their field strengths vanish only modulo \Gl{b4}.
Of course, if the $u_a(x)$
solve \Gl{b4}, then the gauge field matrices
\[ \cb_{\m}(x)=\A \m{I}(x,[u(x)])\, T_I \]
are, at least in a neighborhood of $x$, pure gauges of the form
\beq \cb_{\m}(x)=-G(x)\0 {\6 G^{-1}(x)}{\6x^\m},\qd G(x)=\exp (\ep^I(x)T_I)
\label{puregauge}\eeq
where $\{T_I\}$ denotes a suitable matrix representation of $\cg$
($T_I$ has constant entries). But generally the entries of
$G(x)$ and $G^{-1}(x)$ are not local functions of the form $f(x,[u])$
since \Gl{puregauge} holds only for solutions of \Gl{b4} but not
for arbitrary functions $u_a$.\msk

\noindent{\it Examples:}\\
We consider again the simplest case, namely $N_v=1$. In this case
a `standard form' of the generators of $\cg=sl(2)$ is given by
\beq \L_{-1}=-\0 \6{\6v},\qd \L_{0}=-v\, \0 \6{\6v},\qd
\L_{1}=-v^2\, \0 \6{\6v}\label{g6}\eeq
which satisfy \Gl{sl2}. The corresponding form of the
gauge fields $A_\m$ is
\beq A_\m=\A \m{-1}(x,[u])+\A \m{0}(x,[u])\, v+\A \m{1}(x,[u])\, v^2
\label{g7}\eeq
and the gauge transformations \Gl{g4} are M\"obius transformations given by
\beq \4v=\0 {\a(x,[u])\,  v+\be(x,[u])}{\g(x,[u])\, v+\de(x,[u])},\qd
\a \de-\be\g\in\{1,-1\}\label{26}\eeq
where $\a \de-\be\g\in\{1,-1\}$ can be replaced by
$\a \de-\be\g=1$ without loss of generality if $\a,\ldots,\de$ are
complex\footnote{In the case of real $\a,\ldots,\de$ a restriction
to $\a \de-\be\g=1$ would exclude for instance the gauge transformation
$\4v=-v$.}.
The corresponding finite transformations of the
$\A \m{I}(x,[u])$ under the gauge transformations \Gl{26}
are conveniently written in matrix form.
To this end we use the following matrix representation of \Gl{sl2}
\beq \si_{-1}=\left(\ba{cc}0&1\\0&0\ea\right),\qd
\si_{0}=\left(\ba{cc}\s0 12&0\\0&-\s0 12\ea\right),\qd
\si_{1}=\left(\ba{cc}0&0\\-1&0\ea\right)\label{29}\eeq
and define the matrices
\beq \ca_\m=\sum_{I=-1}^1\A \m{I}\si_I=\left(\ba{cc}\s0 12\A \m{0}&\A \m{-1}\\
-\A \m{1}&-\s0 12\A \m{0}\ea\right),\qd
\cM=\left(\ba{cc}\a&\be\\ \g&\de\ea\right).\label{30}\eeq
The transformed components $\tA \m{I}$ are related to the original components
via
\beq \4\ca_\m=\sum_{I=-1}^1\tA \m{I}\si_I=\cM\, (\ca_\m-\6_\m)\,
\cM^{-1}.\label{31}\eeq
Examples of BTs which are of the form \Gl{g7} are given by
\Gl{bmiu} and \Gl{bkdv}. They also provide an example of two BTs which are
related by a gauge transformation
\Gl{26}. Namely denoting the function $v$ which occurs in \Gl{bmiu}
by $\4v$ and identifying $u=w_x$, the gauge transformation
which relates \Gl{bmiu} and \Gl{bkdv} is given by
\beq \4v=\s0 12\, (w-v).\label{g12}\eeq
Examples for rigid gauge transformations \Gl{13b} which cast the
BTs given in examples a), b) and e) of section \ref{bt} in the
form \Gl{g7} are respectively given by
\beq a)\qd z=\tan \0 v4,\qd b)\qd z=\exp \0 v2,\qd e)\qd z=\tanh \0 v2.
\label{g8}\eeq
\Gl{g8} can also be used to construct closed forms of the gauge
transformations \Gl{g4} for the $sl(2)$-representations arising from
\Gl{table1} in the cases a), b) and e). They are simply given by
\[ \4v(v,x,[u])=\0 {\a\, z(v)+\be}{\g\, z(v)+\de} \]
with $\a,\ldots,\de$ as in \Gl{26} and $z(v)$ as in \Gl{g8}.

\mysection{Construction of B\"acklund transformations
from zero-curvature representations}\label{diff}

The results of the previous sections show that a ZCR $\A \m{I}(x,[u])$
of a system
of PDEs can be used to construct BTs for this system according to
\beq \0 {\6v_i}{\6x^\m}=\A \m{I}(x,[u])\, R_{I i}(v)
\label{d9a}\eeq
where the $R_{I i}$ are obtained
from a representation \Gl{6A} of $\cg$.
The question arises which representations are suited for the
construction of BTs. In this section we discuss nonlinear representations which
are linked with linear representations of $\cg$ and therefore can relate
BTs with inverse scattering methods. This makes them particularly interesting
in the present context.

These representations are constructed by means of
$n\times n$-matrix representations $\{T_I\}$ of $\cg$ which
satisfy\footnote{For the applications we have in mind one has to choose the
representation $\{T_I\}$ such that the structure constants $\f IJK$ in \Gl{c1}
agree with those which occur in \Gl{F2a}.}
\beq [T_I,T_J]=\f IJK T_K.\label{c1}\eeq
A set of differential operators of the form \Gl{6A}
which represent $\cg$ linearly is obviously given by
\beq l_I=-r_{I \a}(\ph)\, \0 \6{\6\ph_\a}, \qd r_{I \a}(\ph)=\T I\a\be \ph_\be
,\qd \a,\be=1,\ldots,n
\label{c2}\eeq
where $\T I\a\be$ are the entries of $T_I$ ($\a$ labels the rows,
$\be$ the columns). BTs of the form \Gl1 are then obtained from a given ZCR by
\beq \0 {\6\ph_\a}{\6x^\m}=\A \m{I}(x,[u])\, \T I\a\be\ph_\be.\label{d1}\eeq
In a more common terminology \Gl{d1}
would not be called a `BT' but rather a `scattering problem' since
it has the form of the linear problems which are used in the inverse
scattering theory \cite{akns}.
Nonlinear representations of $\cg$ can be obtained from the linear
representations \Gl{c2} by means of a set of functions
\beq v_i=v_i(\ph),\qd i=1,\ldots,N_v              \label{d2}\eeq
whose $l_I$-variations can be written completely in terms of the $v_i$
again:
\beq  l_I\, v_i(\ph)=-\T I\a\be\ph_\be\0 {\6v_i}{\6\ph_\a}=-R_{I i}(v).
\label{d3}\eeq
\Gl{d3} represents a nontrivial requirement since \Gl{d2} is not
assumed to be invertible. In particular in general one has $N_v\neq n$.
\Gl{d3} implies immediately that the operators
\beq \L_I=- R_{I i}(v)\, \0 \6{\6v_i}\label{d4}\eeq
represent $\cg$ with the same structure constants which occur in \Gl{c1}:
\beq [T_I,T_J]=\f IJK T_K\ \then\
[l_I,l_J]=\f IJK l_K\ \then\
[\L_I,\L_J]=\f IJK \L_K.\label{c7}\eeq
Thus each choice \Gl{d2} which satisfies \Gl{d3} provides representations
\Gl{6A} of $\cg$ and can be used to construct a BT \Gl1 from a given ZCR
according to \Gl{d9a}. The $\ph_\a$ then may be regarded only as
`auxiliary variables' introduced to construct representations \Gl{6A} of
$\cg$. However it is useful and quite instructive to assume
\beq v_i=v_i(\ph(x))\label{dd3}\eeq
and impose \Gl{d1} on the $\ph_\a$. Namely then \Gl{d9a} follows from \Gl{d1}
since \Gl{dd3}, \Gl{d1} and \Gl{d3} imply
\beann \0 {\6v_i}{\6x^\m}=\0 {\6v_i}{\6\ph_\a}\0 {\6\ph_\a}{\6x^\m}
=\0 {\6v_i}{\6\ph_\a}\A {\m}I\, \T I\a\be \, \ph_\be=
\A \m{I}(x,[u])R_{I i}(v).\eeann
Furthermore \Gl{d3} implies that
the gauge transformations \Gl{g4} of the $v_i$ generated by
the $\L_I$ are induced by the transformations of the $\ph_\a$ which
transform according to a linear representation of the group:
\bea \4v_i&=&\exp \lb g^I(x,[u]) \, \L_I\rb \, v_i
=\exp \lb g^I(x,[u])\, l_I\rb \, v_i(\ph)=v_i(\4\ph),
\label{d5}\\
\4\ph_\a&=&g_\a{}^\be (x,[u])\, \ph_\be,\qd
g(x,[u])=\exp\lb -g^I(x,[u])\, T_I\rb\label{d6}\eea
where the $g_\a{}^\be$ denote the entries of the matrix $g$.
\Gl{d1}, \Gl{dd3} and \Gl{d5} imply
that the finite gauge transformations of the
gauge fields $\A \m{I}$ read in matrix forms
\beq \4\ca_\m(x,[u])=g(x,[u])\lb\ca_\m(x,[u])-\6_\m\rb
g^{-1}(x,[u])\label{d7}\eeq
where $\ca_{\m}(x,[u])$ and $\4\ca_{\m}(x,[u])$ are defined according to
\beq \ca_{\m}(x,[u])=\A \m{I}(x,[u])\, T_I,\qd
   \4\ca_{\m}(x,[u])=\tA \m{I}(x,[u])\, T_I.\label{d8}\eeq
Using these matrices \Gl{z3} takes the form
\beq \6_\m\ca_{\n}(x,[u])-\6_\n\ca_{\m}(x,[u])-[\ca_{\m}(x,[u]),
\ca_{\n}(x,[u])]=0.\label{c18}\eeq
\bsk

\noindent {\it Remark:}\\
As mentioned already in the previous section
the gauge field matrix $\ca_\m(x,[u(x)])$ is
(at least in some neighborhood of $x$) a pure gauge of the form
\Gl{puregauge}, provided $\{u_a(x)\}$ solves \Gl{b4}. Then \Gl{d1} implies
\beq \0 {\6}{\6x^\m}\lb (G^{-1})_\a{}^\be(x)\, \ph_\be(x)\rb=0\qd\then\qd
 \ph_\a(x)=G_\a{}^\be(x)\, \La_\be\label{dd2}\eeq
where $\{\La_\a\}$ is a set of constants and $G(x)$ is the representation
matrix occurring in \Gl{puregauge}. Thus
the functions $v_i(x)$ can be constructed by means of the matrices
$G(x)$ according to \Gl{dd3}.
This uncovers further connections between BTs, ZCRs and inverse
scattering techniques. Notice however that the connection between
$G(x)$ and the corresponding solution of \Gl{b4} is rather involved
since the entries of $G(x)$ are not local functions $f(x,[u])$
as has been pointed out in the previous section.
\bsk

\noindent{\it Examples:}\\
Nontrivial examples of representations \Gl{6A} constructed by
means of the above-described procedure
are obtained for arbitrary $\cg$ by means of `projective coordinates
in $\ph$-space'
\beq v_i=\0 {\ph_i}{\ph_n},\qd i=1,\ldots,n-1.\label{c3}\eeq
Notice that $N_v=n-1$ in this case. One easily checks that \Gl{d3} holds
and reads:
\beq R_{I i}(v)=
\T Iin-\T Inn v_i+\sum_{j=1}^{n-1}(\T Iij v_j-\T Inj v_jv_i),
\qd i=1,\ldots,n-1\label{c5}\eeq
where $\T Inn$ denotes the particular entry of $T_I$ (not its trace).
The generators $\de_I$ constructed by means of \Gl{c5} generalize the
$sl(2)$ representation \Gl{g6} since
\Gl{c5} is a polynomial of degree 2 in the $v_i$---in fact
\Gl{c5} reproduces \Gl{g6} in the case $\cg=sl(2)$ for
the choice $T_I=\si_I$ with $\si_I$ as in \Gl{29}. The gauge transformations
\Gl{g4} arising from \Gl{c5}
generalize the M\"obius transformations \Gl{26} since \Gl{d5} in this case
gives
\beq  \4v_i=\0 {\sum_{j=1}^{n-1}g_i{}^jv_j+g_i{}^n}
               {\sum_{k=1}^{n-1}g_n{}^kv_k+g_n{}^n}
\label{c8}\eeq
where $g_i{}^j$ are the entries of the matrix $g(x,[u])$ occurring
in \Gl{d6}.

For a given ZCR $\A \m{I}(x,[u])$ of a system of PDEs for functions $u_a$
one now can use the functions \Gl{c5} to construct
a BT of the form \Gl1. The result can be written in the form
\beq
\0 {\6v_i}{\6x^\m}=\A \m{I}(x,[u])\, R_{I i}(v)=
(0,\ldots,0,1,0,\ldots,0, -v_i)\, \ca_\m(x,[u])\,
\lb\ba{c}
v_1\\ \vdots\\ v_{n-1}\\1\ea\rb
\label{c16}\eeq
where the 1 in the vector $ (0,\ldots,0,1,0,\ldots, -v_i)$
occurs at the $i$th position.

\mysection{Construction of B\"acklund transformations
for generalized KdV systems}\label{Lax}

This section exemplifies the construction of BTs from ZCRs
for the generalized KdV systems. The latter are defined by
their Lax representation \cite{lax}
\beq \6_tL^{(n)}=[B^{(n,k)},L^{(n)}],\qd n=2,3,\ldots,\qd k=1,2,\ldots
\label{lax}\eeq
where $L^{(n)}$ denotes the $n$th order Lax operator
\beq L^{(n)}
=\6^n+\sum_{i=0}^{n-2}u_{i+1}\6^i,\qd \6=\0 \6{\6x} \label{laxop}\eeq
which depends on $n-1$ functions $u_i(x,t)$. $B^{(n,k)}$ is an operator
of the form
\beq B^{(n,k)}=\6^k+\sum_{i=0}^{k-2}a_i(\{u\})\6^i \label{laxop1}\eeq
where $\{u\}$ denotes collectively the $u_i$ and their
$x$-derivatives. $B^{(n,k)}$ can be constructed for instance by means of
pseudo-differential operators \cite{ds}. \Gl{lax} yields the members of the
$n$th generalized KdV hierarchy in the form
\beq P^{(n,k)}_i([u])=\6_t u_i+p^{(n,k)}_i(\{u\})=0,\qd i=1,\ldots,n-1.
\label{c10}\eeq
\Gl{lax} is the integrability condition for the generalized
Schr\"odinger problem
\beq L^{(n)}\ps=\la\ps,\qd \ps_t=B^{(n,k)}\ps.\label{c11}\eeq
A ZCR of \Gl{c10} can be obtained by writing \Gl{c11}
in matrix form
\beq \6_x \Ps=\cl^{(n)} \Ps, \qd \6_t\Ps=\cb^{(n,k)}\Ps,
\qd \Ps=(\6^{n-1}\ps,\6^{n-2}\ps,\ldots,\6\ps,\ps)^\top\label{c12}\eeq
where $\cl^{(n)}$ denotes the $n\times n$-matrix
\beq \cl^{(n)}=
     \left(\ba{cccccc} 0 &-u_{n-1}   &-u_{n-2} & \ldots & -u_2 & -u_{1}+\la\\
                     1& 0       & 0       & \ldots & 0   & 0  \\
                      0 &1      & 0       & \ldots & 0    & 0  \\
    \vdots &\vdots &\vdots &\vdots &\vdots&\vdots \\
                      0 &0        & 0   & \ldots &   1     & 0
\ea\right)\label{laxmat}\eeq
$\cb^{(n,k)}$ is the $n\times n$-matrix which represents $\6_t$ on
$\Ps$ according to $\ps_t=B^{(n,k)}\ps$. Its construction is
also straightforward but
somewhat involved since one has to use the first equation \Gl{c11}
in order to eliminate partial derivatives $\6^m\ps$ of order $m\geq n$.
The Lax representation \Gl{lax} now indeed takes the form of a
zero-curvature-condition \Gl{c18}:
\beq \6_t\cl^{(n)}-\6_x\cb^{(n,k)}-[\cb^{(n,k)},\cl^{(n)}]=0\label{c19}\eeq
i.e. the matrices $\cl^{(n)}$ and $\cb^{(n,k)}$ are a ZCR of
\Gl{c10} for $\cg=sl(n)$ (both $\cl^{(n)}$ and $\cb^{(n,k)}$ are traceless).
Using the defining representation $\{T_I\}$ of $sl(n)$ (traceless
$n\times n$-matrices) \Gl{c5} thus allows to construct BTs with
$N_v=N_u=n-1$. They are obtained from \Gl{c16} with
\beq \ca_1=\cb^{(n,k)},\qd \ca_2=\cl^{(n)}.\label{c13}\eeq
The `space part' of these BTs reads explicitly
\beq \0 {\6v_1}{\6x}=\la-u_1-\sum_{i=2}^{n-1}u_{n+1-i}v_i-v_1v_{n-1},
\qd 1<i<n:\qd \0 {\6v_i}{\6x}=v_{i-1}-v_iv_{n-1}.\label{c20}\eeq
One can use these BTs for instance to construct BTs
whose space part can be written as a generalized
Miura map of the form
\beq u_i=u_i(\{\4v\}),\qd i=1,\ldots,n-1.\label{c21}\eeq
Namely simple examples of gauge transformations which lead to such
generalized Miura maps are given by
\beq \4v_i=v_i+\sum_{j=0}^{n-2-i}\a_{i,j}\6^ju_{i+j+1},
\qd \a_{i,0}\neq 0\label{c22}\eeq
where $\a_{i,j}$ are constant coefficients. Namely
\Gl{c22} and \Gl{c20} imply
\beq \0 {\6\4v_i}{\6x}=-\a_{i-1,0}\, u_{i}+Y_i(\4v,\{u_{i+1}\},\ldots,
\{u_{n-1}\}),
\qd\a_{0,0}:=1\label{c22a}\eeq
and one easily makes sure that this implies \Gl{c21}.
The BTs obtained in this way relate the generalized
KdV system \Gl{c10} to a similar system of evolution equations
for the $\4v_i$ which has the form
\beq \4Q_i([\4v])=\6_t\4v_i+q_{i}(\{\4v\})=0,\qd i=1,\ldots,n-1.\label{c23}\eeq
This is easily seen combining \Gl{c16}, \Gl{c22} and \Gl{c21}.

One may also look for gauge transformations which allow
to construct auto-BTs for the generalized KdV systems.
To this end it seems reasonable to introduce a potential $w_i$ for
each $u_i$ such that
\beq u_i=\6 w_i\label{c25}\eeq
and look for auto-BTs for the systems of PDEs which arise from
\Gl{c10} for the $w_i$ and are therefore called generalized pKdV systems.
The introduction of the potentials $w_i$ is
suggested by the fact that $w_i$ and $v_i$ have the same dimension $n-i$
(this follows if one assigns dimension 1 to $\6$ and requires that the
operators and equations given above have definite dimension).
Therefore one may hope to find auto-BTs for the generalized pKdV systems
by means of appropriate gauge transformations $\4v_i=\4v_i(v,[w])$
chosen such that $dim(\4v_i)=dim(v_i)$. Simple examples of such
gauge transformations are given by
\beq  \4v_i=\g_iv_i+\sum_{j=0}^{n-1-i}\be_{i,j}\6^jw_{i+j}\label{c26}\eeq
where $\g_i$ and $\be_{i,j}$ are dimensionless constants
(in particular they do not depend on $\la$). Of course \Gl{c26} may be
generalized by allowing for nonlinear contributions with
dimension $n-i$ on the right-hand sides.
\bsk

\noindent{\it Examples}:

(i) The simplest nontrivial KdV system arises for $(n,k)=(2,3)$ and is given
by the KdV equation itself since in this case \Gl{c10} reads (for $u=u_1$)
\beq u_t-\s0 14u_{xxx}-\s0 32uu_x=0\label{c27}\eeq
which takes the form \Gl{kdv} after the rescaling $t\rightarrow -t/4$.
One may check that \Gl{c16} then yields precisely the BT
\Gl{bmiu} after the replacements $v\rightarrow -v$, $\la\rightarrow -\la$.
We know already that the
auto-BT \Gl{bkdv} for the pKdV equation \Gl{pkdv} is related
to \Gl{bmiu} via the gauge transformation \Gl{g12} which indeed is of the
form \Gl{c26}.
\bsk

(ii) The simplest nontrivial generalized KdV system for $n=3$
(Boussinesq hierarchy) is given by $(n,k)=(3,2)$. The
Lax pair, the system of PDEs and its ZCR in matrix form read
\bea
& &L^{(3)}=\6^3+u_2\6+u_1,\qd B^{(3,2)}=\6^2+\s0 23u_2,\label{c28}\\
& &\6_tu_2=-\6^2u_2+2\6u_1,\qd \6_tu_1=\6^2u_1-\s0 23\6^3u_2-\s0 23u_2\6u_2
                                                      ,\label{c28a}\\
& &\cl^{(3)}=
\left(\ba{ccc} 0 &-u_{2}   &-u_{1}+\la\\
                1& 0       & 0        \\
               0 &1        & 0            \ea\right),\\
& &\cb^{(3,2)}=
\left(\ba{ccc} -\s0 13u_2 & \s0 13\6u_2-u_1+\la & \s0 23\6^2u_2-\6u_1\\
               0 &-\s0 13u_2 & \s0 23\6u_2-u_1+\la\\
               1 &0        & \s0 23u_2      \ea\right)
\label{c28b}\eea
and one now constructs easily a BT by means of \Gl{c16}. It reads
\bea \6v_1&=&\la-u_1-u_2v_2-v_1v_2,\qd \6v_2=v_1-(v_2)^2,\nn\\
   \6_tv_1&=&\s0 23\6^2u_2-\6u_1-u_2v_1+v_2(\la-u_1+\s0 13\6u_2)-(v_1)^2,\nn\\
   \6_tv_2&=&\la-u_1+\s0 23\6u_2-u_2v_2-v_1v_2.\label{c28c}\eea
The pKdV system which corresponds to \Gl{c28a} can be written in the form
\beq w_t=\5w_x,\qd \5w_t=-\s0 13w_{xxx}-\s0 23(w_x)^2\label{c29}\eeq
where $w$ and $\5w$ are appropriately defined in terms of the
potentials $w_i$ of the $u_i$:
\beq w=w_2,\qd \5w=2w_1-\6w_2\qd \LRA\qd u_1=\s0 12(\5w_x+w_{xx}),\qd
u_2=w_x.\label{c30}\eeq
Using the notations $v:=\4v_2$, $\5v:=\4v_1$
a gauge transformation \Gl{c26} which yields an auto-BT for \Gl{c29}
is given by
\beq v=3v_2+w,\qd \5v=3v_1+2w_x+\5w.\label{c31}\eeq
In order to check this one may insert \Gl{c31} and its derivatives into
\[ v_t=\5v_x,\qd \5v_t=-\s0 13v_{xxx}-\s0 23(v_x)^2\]
and verify that this yields an identity using \Gl{c28c} and \Gl{c29}.
The auto-BT itself is easily obtained from \Gl{c28c} and \Gl{c31}.
I remark that its space part can be written in a form which is
invariant under $v\lra w$, $\5v\lra -\5w$, $\la\rightarrow -\la$
(such discrete symmetries are typical properties of auto-BTs written in an
appropriate form, see remark at the end of section \ref{bt}):
\beq\ba{l} (v+w)_x=(\5v-\5w)-\s0 13(v-w)^2,\\
 (\5v+\5w)_x=\s0 13(w-v)_{xx}-\s0 23(v-w)(\5v-\5w)+\s0 2{27}(v-w)^3+4\la.
\ea\label{c32}\eeq

\mysection{Construction of zero-curvature representations}\label{meth}

It has been shown how one can construct BTs
for systems of PDEs which have a ZCR.
The question arises how to find a ZCR for a given
system. In special cases the ZCRs may be obtained from the definition
of the respective system of PDEs as in the case of the generalized KdV
systems discussed in the previous section whose ZCRs can be constructed
from their Lax representation.
In fact a ZCR of a given system of PDEs can be viewed as a generalized
Lax representation and one may hope to find new interesting systems of
PDEs by imposing zero-curvature conditions on appropriately chosen
`gauge fields'. This procedure has been applied recently by various
authors, see e.g. \cite{ag,gt,cr}.

In general however the
construction of a ZCR for a given system represents a very nontrivial problem
and of course it may turn out to be impossible. In the following a method
is outlined which allows a systematic search for a ZCR of a given system
after the Lie algebra $\cg$ has been fixed.
The choice of $\cg$ is left as an open problem in the general case.
The outlined method may be regarded as a systematization and generalization
of procedures used for instance in \cite{for}.

Let me first describe the procedure in general and then exemplify it
by applying it to the pKdV equation \Gl{pkdv}.
When looking for a method to determine ZCRs one is faced with the problem
that the zero-curvature conditions \Gl{12} contain arbitrary
functions $r_{\m\n}{}^{I A \r_1\ldots\r_n}(x,[u])$.
This arbitrariness reflects of course the fact that the zero-curvature
conditions do not have to
hold identically in the variables \Gl{b3} but only
modulo \Gl{b4}. The idea is now
to choose a suitable subset $\{w_k\}$
of \Gl{b3} and a corresponding representation of the operators
$\6_\m$ on the $w_k$ such that the $\F \m\n{I}$ must vanish {\it identically}
in these variables. Both the choice
of the $w_k$ and the representation of the $\6_\m$ are obtained from
the system \Gl{b4} itself and encode it. The
outlined method may be formalized using the jet-bundle formalism.
However I found it more instructive to explain it by exemplifying it
for a simple example.

Let me add some remarks before dealing with this example. The choice
of the $\{w_k\}$ eliminates infinitely many variables \Gl{b3} but
the number of remaining variables $w_k$ is still infinite.
However the requirement that the gauge fields $\A \m{I}$ are local functions
means that they actually depend only on a finite subset of
$\{w_k\}$ which is denoted by $\{W_k,\ k=1,\ldots,N_W\}$.
Generally there is a minimal choice of such a finite subset which
can lead to
a nontrivial ZCR. This minimal choice depends on the order of the system
\Gl{b4} resp. on the induced representation of the $\6_\m$ (see
example below). Together with a choice of $\cg$ this converts
\Gl{12} into a well-defined problem in finitely many variables
$W_k$. However we know in advance
that this problem does not have a unique solution. Namely still
one has a `gauge freedom' corresponding to those transformations
\Gl{13} which leave invariant the space of gauge fields
$\A \m{I}(x,W_1,\ldots,W_{N_W})$.
Thus one has to choose a
gauge at some stage of the investigation. However I stress that the
gauge cannot be chosen independently of the particular system \Gl{b4}.
For instance one cannot impose
some standard gauge fixing condition like the Lorentz gauge
$\6^\m \A \m{I}=0$ from the beginning since generally it is not possible
to perform a {\it local} gauge transformation \Gl{13} such that a given
ZCR takes a form satisfying such a standard condition, i.e. it depends
decisively on the system \Gl{b4} which gauge fixing conditions are
compatible with locality.

I remark that the method itself
characterizes the systems of PDEs to which it is applicable.
Namely such systems must allow the choice of a subset
$\{w_k\}$ and a corresponding representation of the $\6_\m$ with
the above-mentioned properties.
Examples for systems of PDEs to which the method is applicable are
Cauchy-Kowalewski systems (see e.g. \cite{her2}).
\bsk

\noindent{\it Example:}\\
The outlined method will now be explained and exemplified
by applying it to the pKdV equation in the form \Gl{pkdv}:
\beq w_t+w_{xxx}+3(w_x)^2=0.\label{pk1}\eeq
First we use
\Gl{pk1} to choose a subset $\{w_k\}$ and establish the corresponding
representation of the $\6_\m\in\{\6_t,\6_x\}$:
By means of \Gl{pk1} we eliminate all derivatives
of $w$ which contain a derivative with respect to $t$, i.e. the
remaining variables
in which \Gl{z3} has to hold identically are given by
\beq w_k:=(\6_x)^kw,\qd k=0,1,2,\ldots.\label{pm7}\eeq
The representation of $\6_t$ and $\6_x$ on these variables
which is induced by \Gl{pk1} reads
\beq \6_xw_k=w_{k+1},\qd
\6_tw_k=-w_{k+3}-3\sum_{m=0}^{k}{k\choose m}
w_{m+1}w_{k+1-m}.\label{pm8}\eeq
Locality of the $\A \m{I}$ requires that
they do not depend
on derivatives of $w$ of higher order than some maximal value $N$, i.e.:
\beq \0 {\6\A \m{I}}{\6w_k}=0\qd \forall\, k>N.
\label{pm9a}\eeq
One easily verifies that \Gl{pm8} and \Gl{pm9a} imply
\bea\lefteqn{\6_t\A 2I-\6_x\A 1I-\f JKI\A 1{J}\A 2{K}
=-\lb w_{N+3}+O(N+1)\rb\0 {\6\A 2I}{\6w_{N}}}\nn\\
& &-w_{N+2}\0 {\6\A 2I}{\6w_{N-1}}
-w_{N+1}\0 {\6\A 2I}{\6w_{N-2}}
-w_{N+1}\0 {\6\A 1I}{\6w_{N}}+O(N)\label{h1}\eea
where $O(N)$ collects terms which do not depend on the $w_k$, $k>N$.
Since \Gl{h1} has to hold identically in the variables $w_k$
the coefficients of $w_{N+3}$, $w_{N+2}$ and $w_{N+1}$
have to vanish separately which gives
\beq \A 2I=f^I(w_0,\ldots,w_{N-2}),\qd
\A 1I=-w_N\0 {\6f^I(w_0,\ldots,w_{N-2})}{\6w_{N-2}}+g^I(w_0,\ldots,w_{N-1}).
\label{h2}\eeq
The smallest value of $N$ which can lead to a nontrivial ZCR obviously is
\beq N=2 \label{n=2}\eeq
which reflects the fact that \Gl{pk1} is of order 3 in the partial
derivatives. The corresponding minimal set $\{W_k\}$ is given by
\beq N=2:\qd \{W_k:\ k=1,2,3\}=\{w,w_x,w_{xx}\}\label{W}\eeq
where we returned to the more familiar notation which denotes
$x$-derivatives by subscripts. We introduce the notation $\3X$ for a vector
with components $X^I$ and $\3X\2\3Y$ for the vector
whose components are given by $\f JKI X^J Y^K$. For the case
\beq \cg=sl(2)\eeq
and a basis satisfying \Gl{sl2} we obtain
\bea & & \3X=(X^{-1},X^0,X^1),\ \3Y=(Y^{-1},Y^0,Y^1)\ \then\ \nn\\
& &\3X\2\3Y=(X^{0}Y^{-1}-X^{-1}Y^{0},
2X^{1}Y^{-1}-2X^{-1}Y^{1},
X^{1}Y^{0}-X^{0}Y^{1}).\label{pm21}\eea
Using this notation we have to determine $\3A_\m(w,w_x,w_{xx})$ such that
\beq 0=\6_t\3A_2-\6_x\3A_1-\3A_1\2\3A_2\label{pm14}\eeq
holds identically in the variables $w_k$ with $\6_x$ and $\6_t$ given in
\Gl{pm8}. In the case $N=2$ \Gl{h2} gives
\beq \3A_2=\3f(w),\qd \3A_1=-w_{xx}\3f{\,}'(w)+\3g (w,w_x)\label{pm15}\eeq
where the prime denotes differentiation with respect to $w$:
\[ X':=\0 {\6X}{\6w}.\]
Inserting \Gl{pm15} into \Gl{pm14} and omitting the arguments
of the functions one obtains
\beq 0=-3(w_x)^2\3f{\,}'+w_{xx}w_x\3f{\,}''-w_{xx}\0 {\6\3g}{\6w_x}-w_x\3g{\,}'
+w_{xx}\3f{\,}'\2\3f-\3g\2\3f.\label{pm16}\eeq
Vanishing of the terms containing $w_{xx}$ requires
\beq \0 {\6\3g}{\6w_x}=w_x\3f{\,}''+\3f{\,}'\2\3f\qd \then\qd
\3g=\s0 12(w_x)^2\3f{\,}''+w_x\3f{\,}'\2\3f+\3k(w)\label{pm17}\eeq
where the function $\3k(w)$ is not determined so far.
If we now insert \Gl{pm17} into \Gl{pm16} all terms containing $w_{xx}$ cancel
and we obtain the following equations by requiring the coefficients
of $(w_x)^n$, $n=3,2,1,0$ to vanish:
\bea 0&= &\3f{\,}''',                                \label{pm18a}\\
     0&= &\3f{\,}'+\s0 12\, \3f{\,}''\2\3f,               \label{pm18c}\\
     \3k{\,}'&= &-(\3f{\,}'\2\3f)\2\3f,              \label{pm18d}\\
     0&= &\3k\2\3f.                                  \label{pm18e}\eea
\Gl{pm18a} immediately implies
\beq\3f=\3a+\3b\, w+\3c\, w^2\label{pm19}\eeq
where $\3a,\3b,\3c$ are constant vectors which must be determined.
\Gl{pm18c} then requires
\beq \3b=\3a\2\3c,\qd 2\3c= \3b\2\3c.\label{pm20}\eeq
Using \Gl{pm20} one may verify that
\beq\3f{\,}'\2\3f=-2\3f+\3d,\qd \3d=2\3a+\3b\2\3a.\label{h3}\eeq
By means of \Gl{h3} one easily makes sure that \Gl{pm18d} gives
\beq \3k=\3l+\3a\2\3d\, w+\s0 12\,\3b\2\3d\, w^2
+\s0 13\,\3c\2\3d\, w^3 \label{pm24}\eeq
where $\3l$ is a constant vector.
$\3a$, $\3b$, $\3c$ and $\3l$ now have to be determined from
\Gl{pm18e} and \Gl{pm20}.
An obvious solution of the second equation \Gl{pm20}
is given by $\3c=0$. However one easily verifies that
\beq \3c=0\ \then\ \3b=0\ \then\
 \3A_2=\3a\ \then\ \ldots\  \then\  \3A_1=\r\3a \label{pm24e}\eeq
with an arbitrary constant $\r$. Thus
the case $\3c=0$ leads to gauge fields $\A \m{I}$ which do not depend
on $w$ and its derivatives and thus to an
uninteresting result. We now consider the case $\3c\neq 0$.
As mentioned above there cannot be a unique nontrivial solution
since we have the freedom of gauge transformations.
Therefore we now choose a gauge. Notice
that the gauge field matrix $\ca_2$ occurring in \Gl{30} is in our case
of the simple form
\bea & & \ca_2=A+B\, w+C\, w^2,\qd
A=\left(\ba{cc}\s0 12a^0&a^{-1}\\ -a^1&-\s0 12a^0\ea\right),\nn\\
& &B=\left(\ba{cc}\s0 12b^0&b^{-1}\\ -b^1&-\s0 12b^0\ea\right),\qd
C=\left(\ba{cc}\s0 12c^0&c^{-1}\\ -c^1&-\s0 12c^0\ea\right)\label{pm25a}\eea
where the entries of $A$, $B$, $C$ are the components of
$\3a$, $\3b$, $\3c$. Since $C$ does not vanish due to $\3c\neq 0$
one can always find a constant matrix $\cM$
with determinant $\pm 1$ such that
\beq\cM\, C\, \cM^{-1}=\left(\ba{cc}0&1\\ -\h&0\ea\right)\label{hh1}\eeq
(since $\h$ is the determinant of $C$ it cannot be fixed by such a
transformation). Therefore
we can assume without loss of generality that $\3c$ is of the form
\beq \3c=(1,0,\h).\label{pp1}\eeq
Notice that the constancy of $\cM$ guarantees
that our original requirement \Gl{n=2} still holds.
We now evaluate \Gl{pm20} explicitly for $\3c$ given by \Gl{pp1}. The
second equation \Gl{pm20} requires
\beq (b^0,2b^{1}-2b^{-1}\h,-b^0\h)=2(1,0,\h)
\qd \LRA\qd b^0=2,\qd b^1=\h=0.\label{pp2}\eeq
Notice that this means
\beq B=\left(\ba{cc}1&b^{-1}\\ 0&-1\ea\right),\qd
C=\left(\ba{cc}0&1\\ 0&0\ea\right)\label{pm25}.\eeq
This result can be simplified by a gauge transformation which
does not change the form of $C$ but simplifies $B$:
\[ \cM=\left(\ba{cc}1&\s0 12b^{-1}\\ 0&1\ea\right)\qd \then\qd
\cM\, B\, \cM^{-1}=\left(\ba{cc}1&0\\ 0&-1\ea\right). \]
Thus we can assume without loss of generality that
\beq \3c=(1,0,0),\qd \3b=(0,2,0).\label{pp3}\eeq
One now straightforwardly solves \Gl{pm18e} and
the first equation \Gl{pm20} with the result
\beq \3a=(\la,0,1),\qd \3l=4\la\, \3a\label{pp4}\eeq
where $\la$ remains undetermined and cannot be fixed by a gauge
transformation. The resulting ZCR reads in vector notation
\beq\ba{l} \3A_2=(\la+w^2,2w,1),\\
    \3A_1=-w_{xx}(2w,2,0)+(4\la w_x+(w_x)^2,0,0)
+(4\la-2w_x)(\la+w^2,2w,1).\ea\label{pp7}\eeq
Let us finally use
\Gl{pp7} to construct a BT of the form \Gl1 with $N_v=1$ using the
representation \Gl{g6} of $sl(2)$. The resulting BT reads
\beq v_x=\la+(v+w)^2,\qd
v_t=(4\la-2w_x)(v+w)^2-2w_{xx}(v+w)+(w_x)^2+2\la\, w_x+4\la^2.\label{pp8}\eeq
Notice that this BT is not among those given in the
section \ref{bt}. Namely both \Gl{bmiu} and \Gl{bkdv} contain
derivatives of $w$ up to order 3, i.e. they are ZCRs with $N=3$
for the pKdV equation (recall that in \Gl{bmiu} one has to
identify $u=w_x$). Solving the `space part' of \Gl{pp8} for $w$ and inserting
the result into the `time part' one may check that the BT \Gl{pp8}
relates the pKdV equation to the following evolution equation for $v$:
\beq v_t=\0 {3\, (v_{xx})^2}{4\, (v_x-\la)}+3(v_x)^2-v_{xxx}.\label{pp9}\eeq
\Gl{pp8} may be viewed as the mother of the
BTs \Gl{bmiu} and \Gl{bkdv} since both can be obtained from it
by a gauge transformation \Gl{13}. Namely one can easily check that the BT
\Gl{bmiu} which relates the KdV to the MKdV equation arises from \Gl{pp8}
through the gauge transformation
\beq \4v=v+w\label{pp10}\eeq
and the identification $u=w_x$.
The auto-BT \Gl{bkdv} for the pKdV equation arises from \Gl{pp8} through
the gauge transformation
\beq \4v=-2v-w\label{pp11}\eeq
where in \Gl{pp10} and \Gl{pp11} $v$ denotes the function which occurs in
\Gl{pp8}. These gauge transformations depend on $w$ and
change the value of $N$ from 2 to 3.

Notice that for $N=2$ and $\cg=sl(2)$ we have obtained a ZCR
of the pKdV equation which
is unique up to gauge transformations
which do not change this value of $N$. This uniqueness
gets lost for higher values of $N$ since for each odd value of $N$
(and each choice of $\cg$) there
is among others a nontrivial solution of
\Gl{pm14} which can be written in the form
\beq N=2m+1:\qd  \3A_1=(T(w_k),0,0,\ldots)\qd \3A_2=(Q(w_k),0,0,\ldots)
\label{cons}\eeq
and corresponds to a local conservation law of the KdV equation
\beq \6_tQ(w_k)=\6_xT(w_k).\label{cons1}\eeq
This suggests a close relationship of BTs and local conservation laws
in the case $D=2$
and shows that in two dimensions local conservation laws can also be
determined by the method outlined in this section. Since
in \Gl{cons} only the first component of $\A \m{I}$ is non-zero
the Lie algebra of course in this case is actually abelian, i.e.
from this point of view local conservation
laws \Gl{cons1} are abelian ZCRs of a system of PDEs.

\bsk

{\it Acknowledgement:} I thank M. Reuter for fruitful discussions and
imparting some background knowledge in the early stages of this work.

\end{document}